\documentstyle[12pt,aps,amsfonts,axodraw] {revtex}
\tighten
\draft
\widetext
\input epsf
\newcommand{\beq}{\begin{equation}}
\newcommand{\eeq}{\end{equation}}
\newcommand{\beqs}{\begin{eqnarray}}
\newcommand{\eeqs}{\end{eqnarray}}
\newcommand{\lsim}{\mathrel{\raisebox{-.6ex}{$\stackrel{\textstyle<}{\sim}$}}}

\begin{document}

\draft

\baselineskip 5.0mm

\bigskip
\bigskip

\title{Neutrino Masses in Theories with Dynamical Electroweak Symmetry
Breaking}

\vspace{6mm}

\author{
Thomas Appelquist$^{a}$ \thanks{email: thomas.appelquist@yale.edu} \and
Robert Shrock$^{b}$ \thanks{email: robert.shrock@sunysb.edu}}

\vspace{6mm}

\address{a \ Physics Department, Sloane Laboratory \\
Yale University 06520 \\
New Haven, CT }

\address{(b) \ C. N. Yang Institute for Theoretical Physics \\
State University of New York \\
Stony Brook, N. Y. 11794 }

\maketitle

\vspace{10mm}

\begin{abstract}

   We address the problem of accounting for light neutrino masses in theories
with dynamical electroweak symmetry breaking.  We discuss this in the
context of a class of (extended) technicolor (ETC) models and analyze the
full set of Dirac and Majorana masses that arise in such theories.  As a
possible solution, we propose a combination of suppressed Dirac masses and a
seesaw involving dynamically generated $|\Delta L|=2$ condensates of
standard-model singlet, ETC-nonsinglet fermions.  We show how this can be
realized in an explicit ETC model.  An important feature of this proposal is
that, because of the suppression of Dirac neutrino mass terms, a seesaw
yielding realistic neutrino masses does not require superheavy Majorana
masses; indeed, these Majorana masses are typically much smaller than the
largest ETC scale.

\end{abstract}


\vspace{16mm}

\newpage
\pagestyle{plain}
\pagenumbering{arabic}

\section{Introduction}

An understanding of the fermion mass spectrum remains an intriguing challenge
for particle physics.  The standard model (SM) accomodates quark and charged
lepton masses by the mechanism of Yukawa couplings to a postulated Higgs boson,
but this does not provide insight into these masses, especially since it
requires small dimensionless Yukawa couplings for all of the charged fermions
except the top quark, ranging down to $10^{-6} - 10^{-5}$ for the electron and
$u$ and $d$ quarks. The standard model has zero neutrino masses, and hence must
be modified to take account of the increasingly strong evidence for the very
small but non-zero neutrino masses and significant lepton mixing from solar and
atmospheric data \cite{sol,atm}, consistent with the K2K accelerator neutrino
experiment \cite{k2k}.

Since masses for the quarks, charged leptons, and observed neutrinos break the
chiral gauge symmetry of the standard model, an explanation of these masses
necessarily involves a model for electroweak symmetry breaking (EWSB).  One
possibility is dynamical electroweak symmetry breaking driven by a strongly
coupled gauge interaction, associated with an exact gauge symmetry, denoted
generically as technicolor (TC) \cite{tc}-\cite{tcrev}. The EWSB arises from
the condensation of technifermion bilinears. The generation of realistic masses
for the charged leptons and $u$, $d$, $s$, $c$, and $b$ quarks seems attainable
in this framework, via extended technicolor, in particular with slowly running
("walking") technicolor.  Although additional ingredients are very likely
necessary to explain the large top-quark mass, we explore here the possibility
that an ETC model of the above type can yield a plausible explanation for small
neutrino masses.  This is a significant challenge for dynamical EWSB models. As
conventionally formulated, these theories have no very large mass scale
analogous to the grand unification scale $M_{GUT}$ that enters in the seesaw
mechanism \cite{seesaw} yielding a Majorana mass $m_\nu \sim m_D^2/m_R$, where
$m_D$ is a Dirac mass and $m_R \sim M_{GUT}$ is the mass characterizing
electroweak-singlet neutrinos.

Although some previous attempts have been made to study this problem
\cite{ssvz,holdom,at94}, it is important to reconsider it in light of later
theoretical and experimental developments.  Refs. \cite{ssvz,at94} did not
include Majorana neutrino mass terms and instead explored a suppression
mechanism for Dirac neutrino masses. This approach does not, however, yield
enough suppression to agree with current experiments. Here we give a general
treatment including both Dirac and Majorana mass terms.  We show how ETC
theories dynamically produce such Majorana mass terms and associated
condensates, violating lepton number $L$ as $|\Delta L|=2$ \cite{lfv}.  We
propose, as a possible solution for how to get light neutrino masses, a
combination of naturally suppressed Dirac masses and a seesaw involving the
dynamically generated Majorana mass terms.  We show how this proposal can be
realized in an explicit ETC model.

\section{ Neutrino Mass Terms in Extended Technicolor Theories}

We first present a general discussion taking the technicolor gauge group to
be SU($N_{TC})$. The set of technifermions includes, as a subset, one
family, viz., $Q_L = {U \choose D}_L$, $L_{TC,L} = {N \choose E}_L$, $U_R$,
$D_R$, $N_R$, $E_R$ transforming according to the fundamental representation
of SU($N_{TC}$) and the usual representations of $G_{SM} = {\rm SU}(3)
\times {\rm SU}(2)_L \times {\rm U}(1)_Y$ (color and TC indices are usually
suppressed). To satisfy constraints from flavor-changing neutral-current
processes, the ETC vector bosons, which can mediate generation-changing
transitions, must have large masses.  We envision that these arise from
self-breaking of the ETC gauge symmetry, which requires that ETC be a
strongly coupled, chiral gauge theory. The self-breaking occurs in stages,
for example at the three stages $\Lambda_1 \sim 10^3$ TeV, $\Lambda_2 \sim
50$ TeV, and $\Lambda_3 \sim 3$ TeV, corresponding to the $N_{gen}=3$
standard-model fermion generations. Then $N_{ETC}=N_{TC}+N_{gen}$.

A particularly attractive choice for the technicolor group, used in the
explicit model to be studied here, is ${\rm SU}(2)_{TC}$, which has the appeal
that it minimizes the TC contributions to the $S$ parameter \cite{nutev} and
can yield walking behavior, allowing for realistically large quark and charged
lepton masses.  With $N_{gen}=3$, the choice $N_{TC}=2$ corresponds to
$N_{ETC}=5$.  With $N_f = 8$ vectorially coupled technifermions in the
fundamental representation, studies suggest that this SU(2)$_{TC}$ theory could
have an (approximate) infrared fixed point (IRFP) in the confining phase with
spontaneous chiral symmetry breaking but near to the phase transition (as a
function of $N_f$ for fixed $N_{TC}$) beyond which the theory would go over
into a nonabelian Coulomb phase \cite{vals,gap}. This approximate IRFP provides
the walking behavior, enhancing the technifermion condensates that control the
quark and charged lepton masses.  The walking can also enhance the masses of
pseudo-Nambu-Goldstone bosons, but further ingredients are likely needed to
ensure the absence of some massless Nambu-Goldstone bosons.

A rough estimate of the quark and charged lepton masses can be made by
considering a one-loop diagram in which a fermion $f_a$ emits a virtual ETC
gauge boson, going to a virtual technifermion $F$ which reabsorbs the gauge
boson, producing the mass term $m_{f_a} \bar f_{a,L} f_{a,R} + h.c.$ with
\beq
m_{f_a} \sim \frac{g_{_{ETC}}^2 \eta_a N_{TC}\Lambda_{TC}^3}{4\pi^2 M_a^2}
\label{mfa}
\eeq
where $M_a \sim g_{_{ETC}}\Lambda_a$ is the mass of the ETC gauge bosons
that gain mass at scale $\Lambda_a$ and $g_{_{ETC}}$ is the running ETC
gauge coupling evaluated at this scale.  In eq. (\ref{mfa}) $\eta_a$ is a
possible enhancement factor incorporating walking, which can be as large as
$\Lambda_a/f_F$ \cite{wtc,eta}, where $f_F$ is the technicolor pseudoscalar
decay constant (for our purposes we can take $f_L \simeq f_Q \equiv f_F$).
We recall that $\Lambda_{TC}$ is determined by using the relation $m_W^2 =
(g^2/4)(N_c f_Q^2 + f_L^2) \simeq (g^2/4)(N_c+1)f_F^2$, which gives $f_F
\simeq 130$ GeV.  In QCD, $f_\pi = 93$ MeV and $\Lambda_{QCD} \sim 170$ MeV,
so that $\Lambda_{QCD}/f_\pi \sim 2$; using this as a guide to technicolor,
we infer $\Lambda_{TC} \sim 260$ MeV.

Technicolor models in general also have a set of electroweak-singlet neutrinos,
$\chi_R=(\chi_1,...,\chi_{n_s})_R$ \cite{rh}, some technicolored and some
techni-singlets, in addition to the left-handed, weak-isospin-doublet neutrinos
and technineutrinos. The contributions to the total neutrino mass matrix,
generated by condensates arising at the TC and ETC scales, are then of three
types: (i) left-handed Majorana, (ii) Dirac, and (iii) right-handed Majorana.
The left-handed Majorana mass terms, which violate $L$ by two units, take the
form
\beq
\sum_{i,j=1}^{N_{ETC}}[n_{iL}^T C (M_L)_{ij} n_{jL}] + h.c.
\label{mleftgen}
\eeq
where $n_L=(\{\nu_\ell \},\{N\})_L$ includes the electroweak-doublet
left-handed neutrinos for $i,j=1,2,3$ and technineutrinos for
$i,j=4,....N_{ETC}$; and $C=i\gamma_2\gamma_0$. Left-handed Majorana masses
violate the electroweak gauge symmetry, and, for technineutrinos, also the
TC symmetry, which is exact. Thus, $(M_L)_{ij}=0$ for $i$ or $j$ $= 4,....
N_{ETC}$. The Dirac mass terms take the form
\beq
\sum_{a=1}^{N_{ETC}} \sum_{s=1}^{n_s} \bar
n_{aL}(M_D)_{as}\chi_{sR} + h.c.
\label{mdirac}
\eeq
Finally, the Majorana bilinears with SM-singlet neutrinos are
\beq
\sum_{s,s^\prime=1}^{n_s} \chi_{s R}^T C (M_R)_{ss^\prime}\chi_{s^\prime R} \ ,
\label{mroperator}
\eeq
In (\ref{mdirac}) and (\ref{mroperator}) $(M_D)_{as}=0$ and
$(M_R)_{ss^\prime}=0$ for technicolor-noninvariant entries.

The full neutrino mass term is then
\beq
-{\cal L}_m =
 {1 \over 2}(\bar n_L \ \overline{\chi^c}_L)
             \left( \begin{array}{cc}
              M_L & M_D \\
              (M_D)^T & M_R \end{array} \right )\left( \begin{array}{c}
      n^{c}_R \\
      \chi_R \end{array} \right ) + h.c.
\label{mnugeneral}
\eeq
Since $(M_L)^T=M_L$ and $(M_R)^T=M_R$, the full $(N_{ETC}+n_s) \times
(N_{ETC}+n_s)$ neutrino mass matrix $M$ in (\ref{mnugeneral}) is complex
symmetric and can be diagonalized by a unitary transformation $U_\nu^\dagger$
as $M_{diag.}=U_\nu^\dagger M (U_\nu^\dagger)^T$. This yields the neutrino
masses and transformation $U_\nu$ relating the group eigenstates
$\nu_L = (\bar n, \overline{\chi^c})_L^T$ and the corresponding mass
eigenstates $\nu_{m,L}$, according to $\nu_{j,L} = \sum_{k=1}^{N_{ETC}+n_s}
(U_\nu)_{jk} \nu_{k,m,L}$, $1 \le j \le N_{ETC}+n_s$ (the elements
$(U_\nu)_{jk}$ connecting techni-singlet and technicolored neutrinos vanish
identically).  The lepton mixing matrix for the observed neutrinos \cite{ms}
$\nu_{\ell,L} = U \nu_{m,L}$ is then given by
\beq
U_{ik} = \sum_{j=1}^3 (U_{\ell,L})_{ij} (U_\nu)_{jk} \ , \quad 1 \le i \le 3,
\quad 1 \le k \le N_{ETC}+n_s
\label{u}
\eeq
where $U_{1k} \equiv U_{ek}$, etc., and where the diagonalization of the
charged lepton mass matrix is carried out by the bi-unitary transformation
$M_{\ell,diag.} = U_{\ell,L} M_\ell U_{\ell,R}^\dagger$.

\section{Specific Extended Technicolor Model}

We next present an analysis of a specific ETC model based on the gauge group $G
= {\rm SU}(5)_{ETC} \times {\rm SU}(2)_{HC} \times G_{SM}$. One additional
gauge interaction, $SU(2)_{HC}$, where HC denotes hypercolor, has been
introduced along with $SU(5)_{ETC}$ and $G_{SM}$. Both the $SU(2)_{HC}$ and
$SU(5)_{ETC}$ interactions become strong, triggering a sequential breaking
pattern. The fermion content of this model is listed below, where the numbers
indicate the representations under ${\rm SU}(5)_{ETC} \times {\rm SU}(2)_{HC}
\times {\rm SU}(3)_c \times {\rm SU}(2)_L$ and the subscript gives the weak
hypercharge:
\beqs
(5,1,3,2)_{1/3,L} \ , \quad\quad & & (5,1,3,1)_{4/3,R} \ , \quad
(5,1,3,1)_{-2/3,R} \cr\cr
(5,1,1,2)_{-1,L}  \ , \quad\quad & & (5,1,1,1)_{-2,R} \ , \quad
(\overline{10},1,1,1)_{0,R} \ , \cr\cr
& & (10,2,1,1)_{0,R} \ .
\label{minimalfermions}
\eeqs
Thus the fermions include quarks and techniquarks in the representations
$(5,1,3,2)_{1/3,L}$, $(5,1,3,1)_{4/3,R}$, and $(5,1,3,1)_{-2/3,R}$, left-handed
charged leptons and neutrinos and technileptons in $(5,1,1,2)_{-1,L}$, and
right-handed charged leptons and technileptons in $(5,1,1,1)_{-2,R}$, together
with SM-singlet fermions $\psi_{ij,R}$ in the antisymmetric tensor
representation $(\overline{10},1,1,1)_{0,R}$.  The unusual assignment of the SM
singlets makes the $SU(5)_{ETC}$ gauge theory chiral. Finally, in order to
render the theory anomaly-free and to provide interactions to help trigger the
symmetry breaking, one adds the hypercolored fields in the $(10,2,1,1)_{0,R}$,
denoted $\zeta^{ij,\alpha}_R$, where $ij$ and $\alpha$ are ETC and HC indices.
Thus, $n_s=30$.  We label the ETC gauge bosons as $(V^i_j)_\mu$, $1 \le i,j \le
5$.  To fix the convention for the lepton number assigned to $\psi_{ij,R}$, we
take it to be $L=1$ in order that Dirac terms $\bar n_{i,L} \psi_{jk,R}$
conserve lepton number.  The lepton number assigned to the
$\zeta^{ij,\alpha}_R$ fields is also a convention; since they have no Dirac
terms with observed neutrinos, we leave it arbitrary.  We write $\chi_R =
(\psi,\zeta)_R$.

Each of the nonabelian factor groups in $G$ is asymptotically free. There are
no bilinear fermion operators invariant under $G$ and hence there are no bare
fermion mass terms.  The SU(2)$_{HC}$ and U(1)$_{HC}$ interactions and the
SU(2)$_{TC}$ subsector of SU(5)$_{TC}$ are vectorial.  This model has some
features in common with the ETC model, denoted AT94, of \cite{at94}, but has
different gauge groups and fermion content.

We next analyze the stages of symmetry breaking.  We envision that at $E \sim
\Lambda_1 \sim 10^3$ TeV, $\alpha_{_{ETC}}$ is sufficiently large to produce
condensation in the attractive channel $(\overline{10},1,1,1)_{0,R} \times
(\overline{10},1,1,1)_{0,R} \to (5,1,1,1)_0$, breaking ${\rm SU}(5)_{ETC} \to
{\rm SU}(4)_{ETC}$.  In the most attractive channel (MAC) analysis this is a
highly attractive channel, with $\Delta C_2=24/5$, although it is not the MAC
itself.  (The MAC is $(\overline {10},1,1,1)_{0,R} \times (10,2,1,1)_{0,R} \to
(1,2,1,1)$, with $\Delta C_2=36/5$; this is undesired since it would break
SU(2)$_{HC}$.).  The desired condensation channel is nearly as strong and is
just as probable within the uncertainties of MAC analyses.  With no loss of
generality, we take the breaking direction in SU(5)$_{ETC}$ as $i=1$; this
entails the separation of the first generation of quarks and leptons from the
components of SU(5)$_{ETC}$ fields with indices lying in the set
$\{2,3,4,5\}$. With respect to the unbroken ${\rm SU}(4)_{ETC}$, we have the
decomposition $(\overline{10},1,1,1)_{0,R} = (\bar 4,1,1,1)_{0,R} + (\bar
6,1,1,1)_{0,R}$ We denote the $(\bar 4,1,1,1)_{0,R}$ and antisymmetric tensor
representation $(\bar 6,1,1,1)_{0,R}$ as $\alpha_{1i R} \equiv \psi_{1i,R}$ for
$2 \le i \le 5$ and $\xi_{ij,R} \equiv \psi_{ij,R}$ for $2 \le i,j \le 5$.  The
associated SU(5)$_{ETC}$-breaking, SU(4)$_{ETC}$-invariant condensate is then
\beq
\langle \epsilon_{1 i j k \ell} \xi^{ij T}_R C \xi^{k \ell}_R \rangle =
4\langle \xi^{23 T}_R C \xi^{45}_R - \xi^{24 T}_R C \xi^{35}_R +
\xi^{25 T}_R C \xi^{34}_R \rangle \ .
\label{xixi}
\eeq
This condensate and the resultant dynamical Majorana mass terms for the six
components of $\xi$ in eq. (\ref{xixi}) violate total lepton number as $|\Delta
L|=2$.  The dynamical formation of Majorana mass terms and violation of total
lepton number is an important feature of these models, providing a
necessary ingredient for a (dynamical) seesaw mechanism \cite{inst}.

At lower scales, depending on relative strengths of couplings, different
symmetry-breaking sequences occur.  One plausible sequence, denoted $G_a$, is
as follows: at $\Lambda_2 \sim 10^2$ TeV, SU(4)$_{ETC}$ and SU(2)$_{HC}$
couplings are sufficiently large to lead together to the condensation
$(4,2,1,1)_{0,R} \times (6,2,1,1)_{0,R} \to (\bar 4,1,1,1)$, breaking ${\rm
SU}(4)_{ETC} \to {\rm SU}(3)_{ETC}$ \cite{at94}.  This condensate is
\beq
\langle \epsilon_{\alpha\beta}\epsilon_{i2jk \ell}\zeta^{ij,\alpha \ T}_R C
\zeta^{k \ell,\beta}_R \rangle = 4\langle \epsilon_{\alpha\beta}(
\zeta^{13,\alpha \ T}_R C \zeta^{45,\beta}_R -
\zeta^{14,\alpha \ T}_R C \zeta^{35,\beta}_R +
        \zeta^{15,\alpha \ T}_R C \zeta^{34,\beta}_R ) \rangle \ ,
\label{zetacondensate}
\eeq
and the twelve $\zeta^{ij,\alpha}_R$ fields in this condensate gain masses
$\sim \Lambda_2$. Both the SU(4)$_{ETC}$ and SU(2)$_{HC}$ interactions are
strongly attractive in this channel, together making the channel an example
of the big-MAC of Ref. \cite{at94}. The fact that the neutrino-like fields
$\alpha_{1i,R}$ transform as a $\bar 4$ of SU(4)$_{ETC}$, while the
left-handed neutrinos and technineutrinos transform as a 4, will lead to a
strong suppression of relevant entries in the Dirac submatrix $M_D$
\cite{ssvz,at94}.

In the $G_a$ symmetry-breaking sequence, at the lowest ETC scale, $\Lambda_3
\sim 3$ TeV, the $(3,2,1,1)_{0,R}$, $\zeta^{2j,\alpha}_R$, $j=3,4,5$, from the
$(6,2,1,1)_{0,R}$ is assumed to condense as $(3,2,1,1)_{0,R} \times
(3,2,1,1)_{0,R} \to (\bar 3,1,1,1)$, breaking ${\rm SU}(3)_{ETC} \to {\rm
SU}(2)_{TC}$ \cite{at94}.  The condensate is $\langle
\epsilon_{\alpha\beta}\zeta^{24,\alpha \ T}_R C \zeta^{25,\beta}_R \rangle$.
This breaking again involves the combination of attractive ETC and HC
interactions \cite{at94}.  Further, we expect that at a scale $\sim \Lambda_3$
the HC interaction produces the condensate $\langle \epsilon_{\alpha\beta}
\zeta^{12,\alpha \ T}_R C \zeta^{23,\beta}_R\rangle$.  Thus, just as the six
$\xi_{ij,R}$ condense out of the theory at energies below $\Lambda_1$, all of
the 20 fields $\zeta^{ij,\alpha}$ in the $(10,2,1,1)_{0,R}$ have condensed out
of the effective theory at energies below $\Lambda_3$.  Since one may assign
lepton number zero to $\zeta^{ij,\alpha}_R$, condensates of the form $\langle
\zeta_R^T C \zeta_R \rangle$ do not necessarily violate total lepton number in
this model.

A different sequence of condensations, denoted $G_b$, can occur if the
SU(2)$_{HC}$ coupling is somewhat smaller.  At a scale $\Lambda_{BHC} \lsim
\Lambda_1$ (BHC = broken HC), the SU(4)$_{ETC}$ interaction produces a
condensation in the channel $(6,2,1,1)_{0,R} \times (6,2,1,1)_{0,R} \to
(1,3,1,1)_0$.  With respect to ETC, this channel has $\Delta C_2 = 5$ and is
hence slightly more attractive than the initial condensation (\ref{xixi})
with $\Delta C_2 = 24/5$, but it can to occur at the somewhat lower scale
$\Lambda_{BHC}$ because it is repulsive with respect to hypercolor. With no
loss of generality, one can orient SU(2)$_{HC}$ axes so that the condensate
is
\beq
\langle \epsilon_{1 i j k \ell} \zeta^{ij,1 \ T}_R C
\zeta^{k \ell,2}_R \rangle + (1 \leftrightarrow 2) \ .
\label{6x6}
\eeq
Since this is an adjoint representation of hypercolor, it breaks ${\rm
SU}(2)_{HC} \to {\rm U}(1)_{HC}$.  We let $\alpha=1,2$ correspond to
$Q_{HC}=\pm 1$ under the U(1)$_{HC}$.  This gives dynamical masses $\sim
\Lambda_{BHC}$ to the twelve $\zeta^{ij,\alpha}_R$ fields involved.

At a lower scale, $\Lambda_{23}$, in the $G_b$ sequence, we envision that a
combination of the SU(4)$_{ETC}$ and U(1)$_{HC}$ attractive interactions
produces the condensation $4 \times 4 \to 6$ with condensate $\langle
\epsilon_{\alpha \beta} \zeta^{12,\alpha \ T}_R C \zeta^{13,\beta}_R \rangle$,
which then breaks ${\rm SU}(4)_{ETC} \to {\rm SU}(2)_{ETC}$ and is
U(1)$_{HC}$-invariant.  Thus, the sequence $G_b$ has only two ETC breaking
scales, $\Lambda_1$ and $\Lambda_{23}$; additional ingredients are needed to
obtain the requisite range of SM fermion masses.  Here we take $\Lambda_{23}
\sim 10$ TeV. Although there is a residual U(1)$_{HC}$ gauge interaction in
these models, its effects are shielded since it does not couple directly to SM
particles.  Finally, for both $G_a$ and $G_b$, at the still lower scale
$\Lambda_{TC} \sim f_F$, technifermion condensation takes place, breaking ${\rm
SU}(2)_L \times {\rm U}(1)_Y \to {\rm U}(1)_{em}$.

\section{Calculations and Results}

The mass matrix $M$ of neutrino-like (colorless and electrically neutral)
states in Eq. (\ref{mnugeneral}) has $N_{ETC}=5$ and $n_s=30$.  Since the
hypercolored fields do not form bilinear condensates and resultant mass terms
with hypercolor singlets, $M$ is block-diagonal, comprised of a $15 \times 15$
block $M_{HCS}$ involving hypercolor-singlet neutrinos and a $20 \times 20$
block $M_{HC}$ involving the hypercolored fermions, $M_{HC}$.  The entries in
the matrix $M$ arise as the high-energy physics is integrated out at each stage
of condensation from $\Lambda_1$ down to $\Lambda_{TC}$. Composite operators of
various dimension are formed, with bilinear condensation then leading to the
masses.  The nonzero entries of $M$ arise in two different ways: (i) directly,
as dynamical masses associated with various condensates, and (ii) via loop
diagrams involving dynamical mass insertions on internal fermion lines and, in
most cases, also mixings among ETC gauge bosons on internal lines.  Since the
ETC gauge boson mixing arises at the level of one or more loops, most graphs
for nonzero type-(ii) elements of $M$ arise at the level of at least two-loop
diagrams.  The different origins for the elements of $M$ give rise to quite
different magnitudes for these elements; in particular, there is substantial
suppression of most type-(ii) entries.  This suppression is not primarily due
to the ETC gauge couplings, which are strong, but to the fact that the diagrams
involve ratios of small scales such as $\Lambda_{TC}$ and lower ETC scales to
larger scales such as $\Lambda_1$.  The $20 \times 20$ matrix $M_{HC}$
involving the $(10,2,1,1)_{0,R}$ fermions contains dynamical fermion mass
entries resulting from the hypercolor condensates and has Tr($M_{HC})=0$.

The matrix of primary interest, $M_{HCS}$, is given by the operator product
\beq
-{\cal L}_{HCS} =
 {1 \over 2}(\bar n_L,  \ \overline{\alpha^c}_L, \ \overline{\xi^c}_L)
  \left( \begin{array}{ccc}
  M_L & (M_D)_{\bar n \alpha} & (M_D)_{\bar n \xi} \\
  (M_D)_{\bar n \alpha}^T & (M_R)_{\alpha \alpha} & (M_R)_{\alpha \xi} \\
  (M_D)_{\bar n \xi}^T    & (M_R)_{\alpha \xi}^T  & (M_R)_{\xi \xi}
                              \end{array} \right )\left( \begin{array}{c}
      n^{c}_R \\
      \alpha_R \\
      \xi_R \end{array} \right ) + h.c.
\label{mhcs}
\eeq
The five-component $n_R^c$, the four-component $\alpha_,R$, and the
six-component $\xi_R$ each contain TC singlets as well as nonsinglets.
One of the two Dirac submatrices is
\beq
(M_D)_{\bar n \alpha} =
 \left( \begin{array}{cccc}
  b_{12} & b_{13} & 0 &    0 \\
  b_{22} & b_{23} & 0 &    0 \\
  b_{32} & b_{33} & 0 &    0 \\
  0             & 0             & 0 & c_1  \\
  0             & 0             & -c_1 & 0 \end{array} \right )
\label{MDnualpha}
\eeq
The vanishing entries are zero because of exact technicolor gauge invariance.
The entry $c_1$ represents a dynamical mass directly generated by technicolor
interactions corresponding to $\sum_{i,j=4,5} \epsilon^{ij} \bar n_{i,L}
\alpha_{1j,R}$, so that $|c_1| \sim \Lambda_{TC}$.  Note that this involves the
antisymmetric, $\epsilon^{ij}$, rather than the $\delta^i_j$, contraction of
SU(2)$_{TC}$ indices and thus makes crucial use of the fact that the
technicolor group is SU(2) rather than SU($N$) with $N \ge 3$.

\begin{center}
\begin{picture}(200,100)(0,0)
\ArrowLine(0,20)(30,20)
\ArrowLine(30,20)(70,20)
\ArrowLine(70,20)(110,20)
\ArrowLine(110,20)(140,20)
\PhotonArc(70,20)(40,0,180){4}{8.5}
\Text(70,20)[]{$\times$}
\Text(70,63)[]{$\times$}
\Text(10,10)[]{$\alpha_{1j,R}$}
\Text(50,10)[]{$\alpha_{14,R}$}
\Text(90,10)[]{$n^5_L$}
\Text(120,10)[l]{$n^i_L$}
\Text(40,65)[]{$V_j^4$}
\Text(100,65)[]{$V_5^i$}
\Text(170,40)[]{$+ \ (4 \leftrightarrow 5)$}
\end{picture}
\end{center}

\begin{figure}
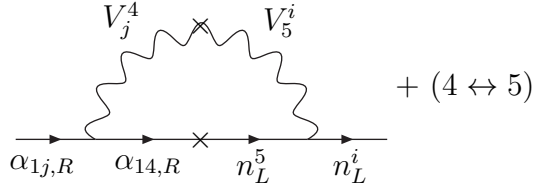

\caption{\footnotesize{Graphs generating $\bar n_{i,L} b_{ij} \alpha_{1j,R}$
for $i=1,2,3$ and $j=2,3$, assuming that the indicated mixings of ETC gauge
bosons occur.}}
\label{alpha-n}
\end{figure}

\begin{center}
\begin{picture}(240,200)(0,0)
\Photon(0,100)(25,100){3}{4.5}
\ArrowArcn(60,100)(35,180,90)
\Text(60,135)[]{$\times$}
\ArrowArcn(60,100)(35,90,0)
\ArrowArcn(60,100)(35,0,270)
\Text(60,65)[]{$\times$}
\ArrowArcn(60,100)(35,270,180)
\Photon(95,100)(120,100){3}{4}
\Text(0,115)[]{$(V_3^4)_\mu$}
\Text(122,115)[]{$(V_5^2)_\nu$}
\Text(25,135)[]{$\zeta^{14,\alpha}_R$}
\Text(20,70)[]{$\zeta^{13,\alpha}_R$}
\Text(105,135)[c]{$\zeta^c_{15,\beta,L}$}
\Text(105,70)[c]{$\zeta^c_{12,\beta,L}$}
\end{picture}
\end{center}

\begin{figure}
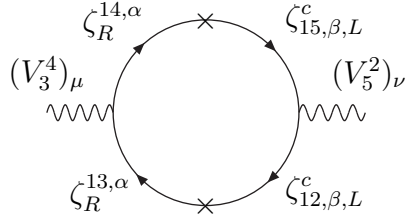

\caption{\footnotesize{One-loop graph contributing to the gauge boson mixing
$V_3^4 \leftrightarrow V_5^2$. The graph with indices 4 and 5 interchanged on
the internal $\zeta$ lines also contributes.}}
\label{vb23}
\end{figure}

In Fig. \ref{alpha-n} we show graphs that could yield the $b_{ij}$'s.  Here the
$\times$ on the fermion line represents the dynamical mass corresponding to a
technicolor condensate.  Each graph requires nondiagonal insertions on
the internal ETC gauge bosons lines. We find that the requisite ETC gauge boson
mixings occur to leading (one-loop) order in the $G_a$ sequence for (i)
$b_{13}$, which involves $V_3^4 \leftrightarrow V_5^1$ and $V_3^5
\leftrightarrow V_4^1$, and (ii) $b_{22}$, which involves $V_2^4
\leftrightarrow V^2_5$; and in the $G_b$ sequence for $b_{23}$ and $b_{32}$,
which involve $V_3^4 \leftrightarrow V_5^2$ and $V_3^5 \leftrightarrow V_4^2$.
For example, for $G_b$ we show in Fig. \ref{vb23} the one-loop graphs
contributing to $V_3^4 \leftrightarrow V_5^2$.  In each respective case, $G_a$
and $G_b$, the other $b_{ij}$'s are produced by higher-loop diagrams.  For
example, starting from Fig. \ref{alpha-n} for $b_{23}$ in case $G_b$, one can
construct diagrams in which the incoming $\alpha_{13,R}$ or the virtual
$\alpha_{14,R}$ or $n^5_L$ emits a virtual $V^k_k$ ETC gauge boson with $k \in
\{1,2,3\}$ which, via mixing, becomes $V^i_2$, which is then absorbed by the
$n^2_L$ to yield an outgoing $n^i_L$, $i=1,3$.  Other similar graphs involving 
a triple ETC gauge-boson vertex along with mixing also contribute in this way.
These generate $b_{13}$ and $b_{33}$
at a level suppressed relative to $b_{23}$.  The $V^k_k \to V^i_2$ mixing
arises generically from loop graphs in which at least one internal fermion line
is a standard-model quark or charged lepton with a mass insertion that is
nondiagonal in generation, incorporating the mixing of the weak eigenstates of
these fermions to form mass eigenstates. The entries $b_{12}$ and $b_{22}$ are
generated in a similar way.

We next estimate the leading $b_{ij}$ entries.  For either breaking
sequence, we denote the ETC gauge boson 2-point function as
\beq
{}^k_n \Pi^i_j(q)_{\mu\lambda} = \int \frac{d^4x}{(2\pi)^4} e^{iq \cdot x}
\langle T \left [ (V^k_n)_\mu(x/2) (V^i_j)_\lambda(-x/2) \right ] \rangle_0 \ .
\eeq
After some manipulations (and Wick rotation), the graph in Fig. \ref{alpha-n}
yields
\beq
g_{ETC}^2 [\bar n_{i,L}(p)\gamma_\mu\gamma_\lambda \alpha_{1j,R}(p)] \
\int \frac{d^4 k}{(2\pi)^4} \
\frac{k^2 \Sigma_{TC}(k) [{}^i_5 \Pi^4_j((p-k)^2)]^{\mu\lambda}}
{(k^2+\Sigma_{TC}(k)^2)^2[(p-k)^2+M_j^2][(p-k)^2+M_i^2]} \ ,
\label{alpha-n-integral_simplified}
\eeq
where $\Sigma_{TC}(k)$ is the dynamical technicolor mass associated with the
transition $\alpha_{14,R} \to n^5_L$.  This mass has the behavior
$\Sigma_{TC}(k) \sim \Lambda_{TC}$ for $k^2 << \Lambda_{TC}^2$, while for $k^2
>> \Lambda_{TC}^2$, (i) $\Sigma_{TC}(k) \sim \Lambda_{TC}^2/k$ for a walking
theory \cite{wtc}, (ii) $\Sigma_{TC}(k) \sim \Lambda_{TC}^3/k^2$ in a QCD-like
theory.  Hence, we need ${}^k_n \Pi^i_j((p-k)^2)_{\mu\lambda}$ only for
$(p-k)^2/\Lambda_1^2 << 1$, since the loop momenta in Fig. \ref{alpha-n} are
cut off far below $\Lambda_1$ (at $\Lambda_3$ for $G_a$ or $\Lambda_{23}$ for
$G_b$). In eq. (\ref{alpha-n-integral_simplified}), $M_j$ denotes the mass
of the ETC gauge boson that picks up mass at $\Lambda_j$.

In the sequence $G_b$, for $q^2 << \Lambda_1^2$, we estimate
\beq
[{}_5^2 \Pi_3^4(q)]_{\mu\lambda} \sim [{}_4^2 \Pi_3^5(q)]_{\mu\lambda} \sim
\frac{g_{ETC}^2\Lambda_{TC}^2}{(2\pi^2)}g_{\mu\lambda} \ .
\label{piv43tov15}
\eeq
where we have assumed a walking behavior of the TC theory up to $\Lambda_{23}$.
For $i,j=2,3$ and $3,2$, adding the other graph with $4 \leftrightarrow 5$
in Fig. \ref{alpha-n}, we find
\beq
|b_{23}| = |b_{32}| \sim \frac{g_{_{ETC}}^4 \Lambda_{TC}^4\Lambda_{23}}
{2\pi^4 M_{23}^4} \sim
\frac{\Lambda_{TC}^4}{2\pi^4 \Lambda_{23}^3}  \quad {\rm for} \ \ G_b \ ,
\label{b23}
\eeq
where we have again assumed the above walking TC behavior.  For sequence $G_a$,
we estimate, using similar methods,
\beq
|b_{13}| \sim \frac{\Lambda_{TC}^2\Lambda_3}{2\pi^4 \Lambda_1^2} \ , \quad
|b_{22}| \sim \frac{\Lambda_{TC}^2\Lambda_3^4}{2\pi^4 \Lambda_2^5} \quad
{\rm for} \ \ G_a \ .
\label{b13b22}
\eeq
With the numerical inputs given above, we get $|b_{23}|=|b_{32}| \sim O(1)$ KeV
for $G_b$ and $|b_{13}| \sim O(1)$ KeV and $|b_{22}| \sim O(10)$ eV for $G_a$.
Because the ETC and TC theories are strongly coupled, these estimates based on
perturbative expansions in powers of $\alpha_{_{ETC}}$ involve an obvious
uncertainty.  For each case, the other $b_{ij}$'s are generated at smaller
levels.  These calculations show how this aspect - suppressed Dirac neutrino
masses - of our proposal are realized in an explicit model.  While the specific
results for the various $b_{ij}$ are dependent on the model and symmetry
breaking pattern, one can infer that this type of suppression can be achieved
in a general class of ETC models where Dirac mass terms are generated in a
similar manner.

The second Dirac submatrix in eq. (\ref{mhcs}) is
\beq
(M_D)_{\bar n \xi} =
 \left( \begin{array}{cccccc}
 d_{1,23} & d_{1,45} & 0 &  0  & 0   & 0   \\
 d_{2,23} & d_{2,45} & 0 &  0  & 0   & 0   \\
 d_{3,23} & d_{3,45} & 0 &  0  & 0   & 0   \\
 0                        & 0  & 0 & c_2 & 0   & c_3 \\
 0                        & 0  &-c_2& 0  &-c_3 & 0  \end{array} \right )
\label{MDnuxi}
\eeq
Again, the zeros are exact and follow from technicolor invariance.  Because the
$\xi$ fields decouple from the theory at scales below $\Lambda_1$, the nonzero
elements of $(M_D)_{\bar n \xi}$ arise indirectly, via loop diagrams and are
highly suppressed.  These elements of $(M_D)_{\bar n \xi}$ have only a small
effect on the neutrino eigenvalues because in the characteristic polynomial
$P(x)$ they occur as corrections to much larger terms involving $\Lambda_1$.

\begin{center}
\begin{picture}(200,100)(0,0)
\ArrowLine(0,20)(30,20)
\ArrowLine(30,20)(70,20)
\ArrowLine(70,20)(110,20)
\ArrowLine(110,20)(140,20)
\PhotonArc(70,20)(40,0,180){4}{8.5}
\Text(70,20)[]{$\times$}
\Text(70,63)[]{$\times$}
\Text(10,10)[]{$\alpha_{13,R}$}
\Text(50,10)[]{$\xi_{43,R}$}
\Text(90,10)[]{$\overline{\xi^c}^{52}_L$}
\Text(120,10)[l]{$\overline{\alpha^c}^{12}_L$}
\Text(40,65)[]{$V_1^4$}
\Text(100,65)[]{$V_5^1$}
\Text(170,40)[]{$+ \ (4 \leftrightarrow 5)$}
\end{picture}
\end{center}

\begin{figure}[h]
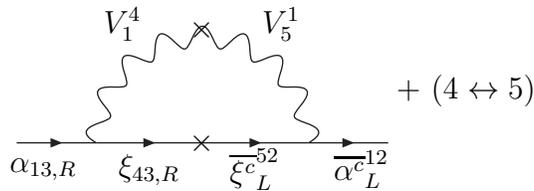

\caption{\footnotesize{Graphs for $\alpha_{12,R}^T C r_{23} \alpha_{13,R}$ in
case $G_b$.}}
\label{alpha-alpha}
\end{figure}

In $M_R$ the $6 \times 6$ submatrix $(M_R)_{\xi \xi}$ has six nonzero entries
that are dynamical mass terms of order $\Lambda_1$ arising directly from the
condensate (\ref{xixi}).  These are important since they are $|\Delta L|=2$
operators, and they, in turn, induce the $(M_R)_{\alpha \alpha}$ Majorana mass
terms which play a central role in the seesaw. Thus the $(M_R)_{\xi \xi}$
entries are the underlying seed for the Majorana mass terms involving the
observed neutrinos. Note that ${\rm Tr}(M_R)=0$.

The submatrix $(M_R)_{\alpha \alpha}$ has the form
\beq
(M_R)_{\alpha \alpha}= \left( \begin{array}{cccc}
 r_{22} & r_{23} & 0  &  0  \\
 r_{23} & r_{33} & 0  &  0  \\
     0  &    0   & 0  &  0  \\
     0  &    0   & 0  &  0  \end{array} \right )
\label{rqmatrix}
\eeq
As before, the zeros are exact and are due to technicolor invariance.  If
the $2 \times 2$ $r_{ij}$ submatrix has maximal rank, this can provide a
seesaw which, in conjunction with the suppression of the Dirac entries
$b_{ij}$ discussed above, can yield adequate suppression of neutrino masses.
The submatrix $r_{ij}$, $2 \le i,j \le 3$, produces this seesaw because
$\alpha_{12,R}$ and $\alpha_{13,R}$ are the electroweak-singlet techni-singlet
neutrinos that remain as part of the low-energy effective field theory at
and below the electroweak scale.

\begin{figure}[h]
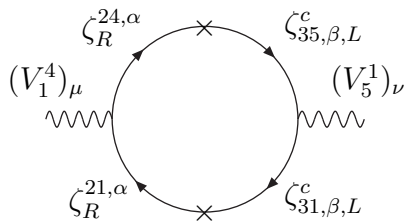

\caption{\footnotesize{One-loop graph for the ETC gauge boson mixing
$V_1^4 \leftrightarrow V^1_5$ in case $G_b$.  The graph with indices 4 and 5
interchanged on the internal $\zeta$ lines also contributes.}}
\label{vr23}
\end{figure}

Consider the sequence $G_b$. In Fig. \ref{alpha-alpha} we show graphs
contributing to $r_{23}$ for this case. These depend on the $V_1^4
\leftrightarrow V^1_5$ ETC gauge mixing produced by the graphs in Fig.
\ref{vr23}.  From these we calculate
\beq
r_{23} \sim \frac{\Lambda_{BHC}^2 \Lambda_{23}^2}{2\pi^4 \Lambda_1^3}
\quad {\rm for} \ \ G_b \ .
\label{r23gb}
\eeq
where here we have assumed a walking behavior of the ETC theory below
$\Lambda_{BHC}$.  The entries $r_{22}$ and $r_{33}$ are generated by
higher-loop diagrams starting from the graphs in Fig. \ref{alpha-alpha} for
$r_{23}$ in a manner similar to that whereby subdominant $b_{ij}$ are generated
starting from Fig. \ref{alpha-n} for $b_{23}$ and $b_{32}$. Numerically, with
the above inputs, $|r_{23}| \sim O(0.1)$ GeV, with smaller values for $r_{ii}$,
$i=2,3$.  For sequence $G_a$ we find that the $r_{ij}$ entries are generated
via higher-loop diagrams analogous to those for $r_{22}$ and $r_{33}$ in
sequence $G_b$ and hence are smaller than eq. (\ref{r23gb}).  In the estimates
to follow we concentrate on the sequence $G_b$ since it yields a
phenomenologically more successful seesaw, although this sequence has only two
ETC breaking scales.

\begin{center}
\begin{picture}(240,200)(0,0)
\Photon(0,100)(25,100){3}{4.5}
\ArrowArcn(60,100)(35,180,90)
\Text(60,135)[]{$\times$}
\ArrowArcn(60,100)(35,90,0)
\ArrowArcn(60,100)(35,0,270)
\Text(60,65)[]{$\times$}
\ArrowArcn(60,100)(35,270,180)
\Photon(95,100)(120,100){3}{4}
\Text(0,115)[]{$(V_1^4)_\mu$}
\Text(122,115)[]{$(V_5^1)_\nu$}
\Text(25,135)[]{$\zeta^{24,\alpha}_R$}
\Text(20,70)[]{$\zeta^{21,\alpha}_R$}
\Text(105,135)[c]{$\zeta^c_{35,\beta,L}$}
\Text(105,70)[c]{$\zeta^c_{31,\beta,L}$}
\end{picture}
\end{center}

In the $4 \times 6$ submatrix $(M_R)_{\alpha \xi}$ the entries are either
exactly zero by technicolor invariance or are nonzero but highly suppressed
because the $\xi$ fields decouple from the effective theory below $\Lambda_1$.
The nonzero entries do not have an important effect on the masses of
neutrino-like states because of the way that they enter in the characteristic
polynomial (similar to the elements of $(M_D)_{\bar n \xi}$).

We next summarize the above discussion from the viewpoint of effective field
theory. At energy scales below $\Lambda_{TC}$, in either the breaking
sequence $G_a$ or $G_b$, the sector of neutrino-like states consists of the
the techni-singlet components $i=1,2,3$ of $n^i_L$ and the techni-singlet
components $\alpha_{1i,R}$, $i=2,3$; other fields have gained masses at
higher scales and have been integrated out. The effective theory comprised
of these degrees of freedom involves bilinear (mass) operators along with a
tower of higher-dimension operators.  The mass operators are either of the
Dirac type (the $b_{ij}$ terms of eq. (\ref{mhcs})) or of the Majorana type
(the $r_{ij}$ of eq. (\ref{rqmatrix})). They form a $5 \times 5$ submatrix
of $M_{HCS}$, and their magnitudes, which depend on the specific breaking
sequence, are $<< \Lambda_{TC}$.

Integrating out the $\alpha_{12,R}$ and $\alpha_{13,R}$ fields then yields the
lowest-scale effective field theory, in which there are three light fermions,
the $n^i_L$. The mass terms in this theory correspond to elements of $M_{L}$,
and there are also higher-dimension operators involving the $n^i_L$. With
respect to the mass terms, this procedure corresponds to a block
diagonalization (``block-seesaw'') of the $5 \times 5$ submatrix of $M_{HCS}$,
keeping only the light, $M_{L}$ matrix. Its dominant terms arise in this
manner; other, smaller entries are generated via higher-loop diagrams involving
higher-dimension operators, for example induced by the exchange and mixing of
ETC gauge bosons. The final step in the effective-field-theory approach is to
diagonalize this $3 \times 3$ matrix, leading to the neutrino mass eigenvalues
and mixing angles.  Equivalently, one can think in terms of diagonalizing the
full $M_{HCS}$-matrix in one fell swoop.

To be specific, we focus on the $G_b$ sequence since it most clearly yields a
seesaw. The largest $M_{L}$ entry is $(M_L)_{23}$ (since $M_L=M_L^T$, we take
$i \le j$.), and other, smaller terms arise from higher dimension operators.
The electroweak-nonsinglet neutrinos are, to very good approximation, linear
combinations of three mass eigenstates, of which the heaviest is $\nu_3$ or
$\nu_2$ and has a mass
\beq
m_{\nu,max} \sim \frac{|b_{23}b_{32}|}{|r_{23}|} \sim
\frac{\Lambda_{TC}^8 \Lambda_1^3}{2\pi^4 \Lambda_{23}^8 \Lambda_{BHC}^2} \ .
\label{mnu3}
\eeq
With the above-mentioned numerical values and $\Lambda_{BHC} \simeq 0.3
\Lambda_1$, we find $m_{\nu,max} \simeq 0.05$ eV, consistent with 
experimental indications \cite{atm} based on a hierarchical spectrum,
in which $m_{\nu,max} \simeq \sqrt{\Delta m^2_{32}}$.  The model naturally
yields large $\nu_\mu-\nu_\tau$ mixing because of the leading off-diagonal
structure of the $b_{ij}$ and $r_{ij}$ with $ij=23$ and $32$. The value of
$|\Delta m^2_{32}|$ depends on details of the model but is on the low side
of the experimental range.  The lightest neutrino mass, $m(\nu_1)$, arises
from the subdominant terms in $M_L$ and is therefore predicted to be
considerably smaller than $m(\nu_i)$, $i=2,3$.  The group eigenstates
involved in these (Majorana) mass eigenstates are $n^c_{i,R}$, $i=1,2,3$ and
$\alpha_{1j,R}$, $j=2,3$. This model thus exhibits our proposed explanation
for light neutrino masses incorporating highly suppressed Dirac neutrino
mass entries, $|\Delta L|=2$ neutrino condensates and associated dynamical
Majorana mass terms, and a resultant seesaw.

The model also yields the following mass eigenvalues and corresponding
eigenvectors for the other neutrino-like states: (i) linear combinations (LC's)
of components of the six $\xi_{ij,R}$ with $2 \le i,j \le 5$ get masses $\sim
\Lambda_1$; (ii) LC's of the $\zeta^{ij,\alpha}_R$ with $2 \le i,j \le 5$ get
masses $\sim \Lambda_{BHC}$; (iii) LC's of the $\zeta^{1j,\alpha}$ with $j=1,2$
get masses $\sim \Lambda_{23}$; (iv) for technicolor nonsinglets, LC's of the
$\zeta^{1j,\alpha}$ with $j=4,5$ and LC's of $n^c_{i,R}$ and $\alpha_{1i,R}$,
with $i=4,5$ get masses $\sim \Lambda_{TC}$; (v) LC's of $\alpha_{1i,R}$ with
$i=2,3$ get masses $\sim r_{23}$.  These masses are (nearly) Dirac.

Not only are the $m_R$ entries responsible for the seesaw not superheavy
masses; they are actually much smaller than the ETC scales $\Lambda_i$. A
generic prediction of ETC models with the proposed seesaw is that some
components of SM-singlet neutrino group eigenstates comprise dominant parts of
mass eigenstates with masses given by the elements in $M_R$ that are involved
in the seesaw (here, $r_{23}$).  A condition to fit current limits on the
emission of massive neutrinos, via lepton mixing, in particle decays would be
that the $|U_{ek}|^2, \ |U_{\mu k}|^2 \lsim 10^{-7}$ for $k > 3$ \cite{rs,pdg},
which can be met while also maintaining sufficiently short lifetimes to satisfy
astrophysical constraints.

\section{Conclusions}

In summary, we have given a general analysis of neutrino masses in the context
of dynamical electroweak symmetry breaking theories, taking account of both
Dirac and Majorana mass terms.  We proposed a possible solution to the problem
of obtaining light neutrino masses in this class of theories.  This solution
involves two main parts: (i) strong suppression of Dirac neutrino masses, and
(ii) dynamical formation of bilinear Majorana neutrino condensates at ETC
scales and resultant Majorana masses violating total lepton number as $|\Delta
L|=2$, and consequently a seesaw mechanism.  We have shown how this proposal
can be realized in an explicit ETC model. While further work is needed to
obtain the detailed structural features needed to fit current indications for
neutrino masses and lepton mixing, we believe that our proposal contains key
ingredients for a solution to this problem in the context of theories with
dynamical electroweak symmetry breaking. An important aspect of this suggestion
is that it does not need any superheavy scale for a viable seesaw; indeed, the
relevant Majorana masses may be much smaller than the highest ETC scale.

This research was partially supported by the grants DE-FG02-92ER-4074 (T.A.)
and NSF-PHY-00-98527 (R.S.).  T.A. thanks B. Dobrescu for helpful comments.

\vfill
\eject


\begin{thebibliography}{99}
\bibitem{lfv}{In general, lepton family number is violated by both Dirac and
Majorana mass terms.}
\bibitem{gap}{In the approximation of a single-gauge-boson exchange, the
critical coupling for condensation $R_1 \times R_2 \to R_c$ is given by the
condition $\frac{3\alpha}{2\pi}\Delta C_2 = 1$, where $\Delta C_2 =
[C_2(R_1)+C_2(R_2)-C_2(R_c)]$, and $C_2(R)$ is the quadratic Casimir invariant.
Instanton contributions are also important \cite{precision}.}
\bibitem{rh}{We write SM-singlet neutrinos as right-handed fields
$\chi_{j,R}$.}
\bibitem{inst}{Since ETC is a chiral gauge theory which is strongly interacting
at its self-breaking scale (here, $\Lambda_1$) and since
$\pi_3({\rm SU}(N))= {\mathbb Z}$, the associated ETC instantons will
generically violate total lepton number $L$.  Because $g_{_{ETC}} \sim O(1)$ at
$\Lambda_1$, these instantons are not suppressed by small
$e^{-16\pi^2/g^2}$ factors, in contrast to the situation at zero temperature
in the weak SU(2) sector.  However, the resultant effective multifermion
operators are of quite high dimension and are thus suppressed at low energies.}
\end{thebibliography}
\end{document}